\documentstyle[preprint,aps,eqsecnum]{revtex}

\newcommand{\be}{\begin{equation}}
\newcommand{\ee}{\end{equation}}
\newcommand{\bea}{\begin{eqnarray}}
\newcommand{\eea}{\end{eqnarray}}
\newcommand{\nn}{\nonumber \\}
\newcommand{\p}{\partial}

\begin{document}

\tightenlines

\draft
\preprint{\small CGPG-97/7-1}

\title{Black Holes with Short Hair}

\author{J.~David Brown$^*$ and Viqar Husain$^\dagger$}

\address
 {$^\dagger$Center for Gravitational Physics
and Geometry,\\
Department of Physics, Pennsylvania State University, \\
 University Park, PA 16802-6300, USA.\\
$^*$Department of Physics and Department of Mathematics,\\
North Carolina State University,\\
 Raleigh, NC 27695-8202 USA.}

\maketitle

\begin{abstract}

We present spherically symmetric black hole solutions for Einstein
gravity coupled to anisotropic matter. We show that these black holes
have arbitrarily short hair, and argue for stability by showing that
they can arise from dynamical collapse. We also show that a recent `no
short hair' theorem does not apply to these solutions.

\end{abstract}

\bigskip
\pacs{Pacs numbers: 04.20.Jb, 04.40.Nr, 04.70.Bw}


\section{Introduction}


It is widely believed that black holes retain only very limited
information about the matter that collapsed to form them. This
information is reflected in the number of parameters characterizing
the black hole solution. For stationary axis--symmetric black holes in
general relativity with no matter coupling, the only such parameters
are the mass $M$ and angular momentum $J$. With coupling to
electromagnetism, the parameters include the electric charge
$Q$. These results are known collectively as the `no hair'
theorems\cite{israel,carter,hawking,robinson,mazur,bunting}.

The three parameters $M$, $J$, and $Q$ are special in that they can be
detected by performing experiments in the asymptotically flat region
of the black hole spacetime.  This is because they are captured by
conserved surface integrals in the asymptotically flat region. They
are not what is referred to as `hair'.  Any parameters in a black hole
spacetime that are not captured by asymptotic surface integrals may be
called `hair'. Thus, no hair theorems imply that all the parameters
characterizing black hole spacetimes are those that are `visible' at
spatial infinity.

For Einstein gravity coupled to matter fields other than
electromagetism, black holes might be parametrized by quantities other
than $M$, $J$, and $Q$. In these cases, the parameters might all be
captured by surface integrals at infinity, leading to an associated
`no hair theorem'. Alternatively, such theories might yield black hole
solutions with hair.

In this paper we describe spherically symmetric and static black hole
solutions that arise from coupling Einstein gravity to an `anisotropic
fluid', which is defined below. For certain special cases this matter
gives the de--Sitter, and Reissner--Nordstrom black holes. For other
cases, the resulting black holes have hair in the sense described
above. In the next section we describe the anisotropic medium and
construct a class of static, spherically summetric solutions to the
coupled Einstein--matter equations of motion. In section III we
specialize to a class of black hole solutions and show that these
black holes have short hair. We also give arguments indicating that
these black holes are stable. Section IV discusses the relationship of
our black hole solutions with the so called `no short hair' conjecture
of Ref.~\cite{nqs}. We conclude in section V with a short discussion.


\section{Solutions with Anisotropic Matter}


Consider a static, spherically symmetric metric of the form
\be
ds^2 = -f(r) dt^2 + \frac{1}{f(r)} dr^2 + r^2 d\theta^2 +
r^2 \sin^2\theta d\phi^2
\label{met}
\ee
The nonvanishing components of the Einstein tensor for this metric are
\bea
G^t_t & = & G^r_r = \frac{ rf' -1 + f}{r^2} \nn
G^\theta_\theta & = & G^\phi_\phi = \frac{rf'' + 2f'}{2r} \ ,
\label{G}
\eea
where the prime denotes a derivative with respect to $r$.  It follows
that any matter source that gives rise to a metric of the form
Eq.~(2.1) must satisfy $T^t_t = T^r_r$ and $T^\theta_\theta =
T^\phi_\phi$. In particular, the ``radial pressure'' $T^r_r$ need not
equal the ``angular pressure" $T^\theta_\theta =
T^\phi_\phi$. Therefore, generically, such a matter source is not
isotropic in the static time slices. What we seek, then, is a medium
that is capable of supporting anisotropic stresses.

A material with the desired properties is the elastic medium
discussed by DeWitt \cite{dewitt}.  We begin with DeWitt's action,
as written in Ref.~\cite{dd},
\be
S[g_{ab},Z^i] = - \int_{\cal M} d^4x \sqrt{-g}\, \rho(Z^i, h_{jk}) \ .
\label{act}
\ee
This action is a functional of the spacetime metric $g_{ab}$ and the
Lagrangian coordinates $Z^i(x)$, $i=1,2,3$. The Lagrangian coordinates
are a set of labels $\zeta^i=Z^i(x)$ that tell which particle passes
through the spacetime point $x$. Thus, $Z^i(x)$ are three scalar fields
whose gradients are orthogonal to the matter world lines.
The action $S[g_{ab},Z^i]$ is the proper volume integral
of the Lagrangian $-\rho$, where $\rho$ is the proper energy density in
the rest frame of the matter. The energy density $\rho$ depends on
$Z^i$ explicitly, and also implicitly through the matter space metric
$h_{ij}$. The matter space metric is defined by
\be
h^{ij} =   (\partial_a Z^i) g^{ab}(\partial_b Z^j)  \ ,
\label{mmet}
\ee
and is interpreted as the metric in the rest frame of the matter. That
is, $h_{ij} d\zeta^i d\zeta^j$ is the square of the proper orthogonal
distance separating particle world lines with Lagrangian coordinates
$\zeta^i$ and $\zeta^i + d\zeta^i$. Using $h_{ij}$, the spacetime
metric $g_{ab}$ may be written in the form
\be
 g_{ab}  = - U_a U_b + h_{ij} (\partial_a Z^i) (\partial_b Z^j)\ ,
\label{3+1}
\ee
where $U^a$ is the fluid velocity, defined as the future pointing unit
vector orthogonal to the gradients $\p_a Z^i$.

The stress--energy--momentum tensor for the elastic medium is
\bea
T_{ab} &\equiv& -\frac{2}{\sqrt{-g}} \frac{\delta
S}{\delta g^{ab}}= -\rho g_{ab}
+ 2 {\p\rho \over\p h^{ij}}(\p_a Z^i)(\p_b Z^j)\nn
&=& \rho U_a U_b + t_{ij} (\p_a Z^i)(\p_b Z^j) \ ,
\label{set}
\eea
where we have used (\ref{3+1}) and defined the matter stress tensor
\be
t_{ij} = \left( 2{\p\rho\over \p h^{ij}} - \rho h_{ij} \right)
= {2\over \sqrt{h}}{\p(\sqrt{h}\rho)\over \p h^{ij}}\ .
\label{mset}
\ee
The matter equations of motion, obtained by varying the action with
respect to $Z^i$, are equivalent to the conservation equations
$\nabla_a T^{ab} = 0$.  When the Einstein equations hold, the
contracted Bianchi identity implies that the matter equations are also
satisfied.

An {\it isotropic} elastic medium is one whose principal pressures
are all equal. Its stress tensor is that of a perfect fluid,
and is obtained by letting $\rho$ depend only on the proper volume of
the Lagrangian coordinate cell $d^3\zeta$ and the number of particles
in that cell. This is realized \cite{dd} by setting
\be
\rho = \rho(N)\ ,
\ee
with the scalar $N$  defined by the ratio of densities
\be
N = \underline{N}(Z^i)/\sqrt{h}\ ,
\ee
where $\underline{N}(\zeta^i)$ is the number of particles per unit
coordinate cell $d^3\zeta$. Thus, $N$ is the proper number density,
i.e., the number of particles per unit proper volume in the material
rest frame. With these definitions, the stress--energy--momentum
tensor (\ref{set}) assumes the familiar form: Using the identity
\be
{\p\rho(N) \over\p h^{ij}}= {N\over 2} {\p\rho \over\p N}h_{ij}\ ,
\ee
and (\ref{3+1}), the tensor (\ref{set}) becomes
\be
T_{ab} = \rho U_aU_b + \left( N {\p\rho \over\p N}-\rho \right)
(g_{ab} + U_a U_b)\ .
\ee
Thus, the matter action (\ref{act}) with $\rho=\rho(N)$ is the action
of a perfect fluid with pressure $P=N (\p\rho/\p N)-\rho$. This
definition of pressure can be written as $d(\rho/N) = - P d(1/N)$, which
is the first law of thermodynamics applied to a cell with a fixed
number of particles.

An {\it anisotropic} elastic medium is one whose principal pressures
are not all equal. If two of the three principal pressures are equal,
the fluid has planar symmetry.  The basic idea for introducing
anisotropy is to perform a `$2+1$' decomposition of the matter space
metric $h_{ij}$, and treat the plane differently from the line. This
may be done as follows. Let us view the Lagrangian coordinates
$\zeta^i$ as points in a manifold ${\cal S}$, which is the space of
particle flow lines, and introduce a foliation of ${\cal S}$ by a
function $\varphi(\zeta^i)$ with $\partial_i\varphi \neq 0$.  Now set
$\rho= \rho(n,m)$, where $n$ is the particle number per unit proper
area in the $\varphi={\rm const}$ slices and $m$ is the particle
number per unit proper length in the direction orthogonal to these
slices. This introduces the anisotropy.  For definiteness and
simplicity, let us choose $\varphi(\zeta^i) = \zeta^1$.  Then $n =
{\underline n}/\sqrt{\sigma}$ where $\underline n(\zeta^i)$ is the
number of particles per unit coordinate area $d^2\zeta =
d\zeta^2\wedge d\zeta^3$, and $\sigma = h_{22}h_{33} - (h_{23})^2$ is
the determinant of the metric in the $\zeta^1 = {\rm const}$
slices. Similarly, $m = {\underline m}/\alpha$ where ${\underline m}$
is the number of particles per unit coordinate length $d\zeta^1$ and
$\alpha = \sqrt{h/\sigma}$ is the `lapse function' that measures
proper length in the direction orthogonal to $\zeta^1 = {\rm const}$.

We thus have the following `$2+1$' form for $h_{ij}$:
\be
h_{ij}= \left( \matrix{ \alpha^2 + \beta^\mu\beta_\mu & \beta_\nu \cr
                           \beta_\mu    & \sigma_{\mu\nu} }
        \right)\ ,
\ee
where $\beta^\mu$ is the `shift vector' and the indices $\mu$,
$\nu$ take the values $1$, $2$.
Equivalently, we have $h_{ij} = u_iu_j + \sigma_{ij}$, where $u_i =
\alpha\delta_i^1$ is the unit vector orthogonal to the $\zeta^1
=\rm{const.}$ surfaces. The corresponding spacetime tensors are
\be
u_a = u_i\p_aZ^i = \alpha\p_aZ^1\ ,  \ \ \ \
\sigma_{ab} = (h_{ij} - u_iu_j)(\partial_a Z^i)(\partial_b Z^j)\ ,
\ee
which represent, respectively, the unit normal and metric of the
isotropic 2-surfaces in the rest frame of the matter.

The stress tensor for the anisotropic medium, with $\rho(n,m)$, is
obtained by computing $(\p\rho/\p h^{ij})$ using the above definitions,
and substituting into (\ref{mset}). We find
\be
{\p \rho(n,m)\over \p h^{ij} } =
{n\over 2}{\p \rho\over \p n}\sigma_{ij}
+ {m\over 2}{\p \rho\over \p m}u_i u_j
\ee
which gives
\be
t_{ij} = \left( n\frac{\p\rho}{\p n} - \rho \right) \sigma_{ij}
+ \left( m\frac{\p\rho}{\p m} - \rho  \right) u_i u_j \ .
\ee
Finally, the full stress--energy--momentum tensor for the
anisotropic matter is
\be
T_{ab} = \rho U_a U_b + \left( n\frac{\p\rho}{\p n} - \rho\right)
\sigma_{ab}
+ \left( m\frac{\p\rho}{\p m} - \rho\right) u_a u_b
\ .
\label{aset}
\ee
We note, as required, that the pressure in the 2-surfaces is
$n(\partial\rho/\partial n)- \rho$, which differs from the pressure
$m(\partial\rho/\partial m) - \rho$ in the direction orthogonal to
2-surfaces.

Our goal now is to find an equation of state, $\rho = \rho(n,m)$, that
will yield the metric (\ref{met}) as a solution of the Einstein and
matter equations of motion, {\it for arbitrary} $f(r)$.  The Einstein
tensor (\ref{G}) shows that the spheres $t={\rm const}$, $r={\rm
const}$ are isotropic. Thus, we seek a solution where $u_a$ is
radial. In particular, let $Z^1(x) = r$, $Z^2(x) = \theta$, and
$Z^3(x) = \phi$ so that $u_a = \alpha \delta_a^r$. From (\ref{aset})
the condition $T_t^t = T_r^r$ now shows that the equation of state
must be independent of $m$, so $\rho = \rho(n)$. In this case we have
\bea
T^t_t & = & T^r_r = -\rho \\
T^\theta_\theta & = & T^\phi_\phi = n\frac{\p\rho}{\p n} -
\rho  \ ,
\eea
and the Einstein equations yield (with $G = 1$)
\bea
   \frac{ rf'- 1 + f}{r^2}  & = & -8\pi\rho \label{EE1}\\
 \frac{rf'' + 2f'}{2r}
 & = & 8\pi \left(n\frac{\partial\rho}{\partial n} - \rho\right) \ .
\label{EE2}
\eea
Using the relationship $(rf' -1 +f)' = rf'' + 2f'$ between the
numerators of the left--hand sides of these equations, we find
\be
\frac{(-r^2\rho)'}{2r} = n\frac{\partial\rho}{\partial n} - \rho\ .
\ee
This reduces to $n'/n = -2/r$, and implies
\be
n = \frac{K}{r^2} \ ,
\ee
where $K$ is a positive dimensionless constant. Now define the function
$F(n)$ by
\be
F(n) \equiv F(K/r^2) \equiv f(r) \ .
\ee
Then the Einstein equation (\ref{EE1}) gives
\be
\rho(n) = \frac{n}{8\pi K} \left( 1 - F + 2n\frac{\p F}{\p n}
\right) \ .
\label{eos}
\ee
This determines the equation of
state $\rho(n)$ in terms of the {\it arbitrary function} $f(r)$ that
appears in the metric (\ref{met}).

Note that the correspondence between $f(r)$ and $\rho(n)$ is not
one--to--one.
One can see this by solving Eq.~(\ref{eos}) for $F(n)$:
\be
F(n) = 1 + 4\pi K \sqrt{n} \int dn \, n^{-5/2} \rho(n) \ .
\label{F(n)}
\ee
The integral over $n$
yields a term $4\pi K K'\sqrt{n}$ which is proportional to the
integration constant $K'$. This term corresponds to $-2M/r$ in
the metric function $f(r)$, where $M = -2\pi K^{3/2}K'$.

Let us reexpress and summarize these results. Consider an elastic
medium (\ref{act}--\ref{mmet}) with equation of state $\rho = \rho(n)$,
where $n = {\underline n}/\sqrt{\sigma}$, ${\underline n} = K\sin
Z^2$, and $\sigma = h_{22} h_{33} - (h_{23})^2$. A family of solutions of
the Einstein and matter equations of motion is $Z^1(x) = r$, $Z^2(x) =
\theta$, $Z^3(x) = \phi$, and $g_{ab}(x)$, where $g_{ab}(x)$ is given
by (\ref{met}). The function $f(r)$ that appears in the metric is defined
by $f(r) = F(K/r^2)$, where $F(n)$ is
determined by the equation of state through Eq.~(\ref{F(n)}).
This family of solutions is parametrized by an arbitrary constant $M$
that appears in a term $-2M/r$ in the metric function $f(r)$.
Equation (\ref{eos}) shows that, for a suitable equation
of state $\rho = \rho(n)$, any metric of the form (\ref{met}) can be
obtained as a solution of the Einstein and matter
equations of motion.

The anisotropic elastic medium described here satisfies the (weak,
dominant, or strong) energy conditions only if restrictions are placed
on the equation of state. For example, for the weak energy
condition the proper energy density must be nonnegative, $\rho\geq 0$,
and the principle stresses must be greater than or equal to $-\rho$.
This later requirement implies $\partial\rho/\partial n \geq 0$. The
inequalities $\rho\geq 0$ and $\partial\rho/\partial n \geq 0$ can be
translated into restrictions on the function $F(n)$, which in turn
restrict the form of the metric function $f(r)$.


\section{Black holes with short hair}


In the previous section we showed that the general spherically
symmetric metric (\ref{met}) arises from an anisotropic matter
source. However all such metrics are not physical because, as
mentioned above, the matter does not in general satisfy physical
energy conditions. In this section we consider a class of black hole
solutions of the Einstein--anisotropic fluid theory that are
physical. We show that these black holes have arbitrarily short hair,
and argue that they are stable under scalar perturbations.  The
stability argument relies on the observation that these black holes
can arise from spherically symmetric gravitational collapse.

Consider the equation of state
\be
 \rho(n) = Cn^{k+1}
\label{nk+1}
\ee
for the anisotropic fluid, where $C$ and $k$ are constants. From
Eq. (\ref{aset}), the stress--energy--momentum tensor is
\be
T_{ab} = \rho U_aU_b -\rho u_au_b + P \sigma_{ab}\ ,
\label{aset2}
\ee
where
\be
 P \equiv  n{\p\rho \over \p n}-\rho = k\rho
\ee
defines the principal pressures in the $\theta-\phi$ surfaces.

$T_{ab}$ is a type I stress tensor, so the weak energy condition
requires $\rho\ge 0$ and $P\ge -\rho$. This clearly holds for all
$C\geq 0$, $k\geq -1$.  The strong energy condition requires, in
addition to the weak energy condition, that the sum of the energy
density $\rho$ and the principal pressures should be non--negative. In
our case this implies $P\geq 0$, so the strong energy condition holds
for all $C\geq 0$, $k\geq 0$.  The dominant energy condition requires,
in addition to the weak energy condition, that energy fluxes should
never be spacelike. In our case this leads to $\rho\geq P$. Therefore
the dominant energy condition holds for $C\geq 0$, $-1 \leq k \leq 1$.

{}From Eq. (\ref{F(n)}) we see that the metric resulting from
the equation of state $\rho(n) = Cn^{k+1} $ is (\ref{met})
with
\be
f(r) = \left( 1- {2M\over r} + {Q^{2k}\over r^{2k}}\right)\ ,
\ \ \ \ \ (k\ne 1/2) \ .
\label{sh}
\ee Here, we have chosen $C>0$ and defined $Q^{2k} = 8\pi C K^{k+1}/(2k-1)$.
The energy density and pressures are
\be
\rho={Q^{2k}(2k-1)\over 8\pi r^{2k+2}}\ ,\ \ \ \ \
P=k\ {Q^{2k}(2k-1)\over 8\pi r^{2k+2}}\ .
\label{rhoP}
\ee
Notice
that for $k=1$ this solution is identical to the Reissner--Nordstrom
metric. For our purposes, the more interesting situation occurs for
$k>1$. In that case the parameter $Q$ is {\it not} captured by a
surface integral at spatial infinity due to the rapid fall off
$r^{-2k}$ of the corresponding term in $f(r)$. Therefore, when $k>1$,
$Q$ qualifies as hair. Since we can take $k$ to be arbitrarily large,
{\it the hair $Q$ can be arbitrarily short}. This means that the
dependence of the metric on $Q$ may be confined to an arbitrarily
small neighbourhood of the curvature singularity at $r=0$.
As an example, consider the limit $Q\ll M$. In this case the outer
event horizon is at $r_+ \approx 2M$ and the inner (Cauchy) horizon
is at $r_- \approx Q(Q/2M)^{1/(2k-1)} + 2M(Q/2M)^{4k/(2k-1)}/(2k-1)$.
The ratio of the energy densities
at the event and inner horizons is $\rho(r_+)/\rho(r_-) = (r_-/r_+)^{2k+2}
\approx (Q/2M)^{2k(2k+2)/(2k-1)}$. This ratio can be made arbitrarily
small by decreasing $Q/2M$ or by increasing $k$.
Thus the hair sprouting from the event horizon can be arbitrarily short.
 We emphasize that the anisotropic fluid that gives rise to
these short--hair solutions satisfies both the weak and strong energy
conditions.

We now consider the issue of stability for the black holes with short
hair. In particular, we present evidence that the black holes
(\ref{sh}) are stable under the spherically symmetric
perturbations
\be
f(r)\rightarrow f(r) + \delta f(r,t)\ ,\ \ \
\rho(r)\rightarrow \rho(r) + \delta \rho(r,t)\ , \ \ \
P(r)\rightarrow P(r) + \delta P(r,t)
\label{Del}
\ee
about the static solutions. As discussed below, these are not the most
general possible spherically symmetric perturbations.

Stability of a static or stationary black hole is studied by
performing a standard but tedious analysis to obtain the equations
governing metric and fluid perturbations to linearized order. The
black hole is stable if initial metric perturbations do not grow
without bound. Alternatively, one can attempt to show that the black
hole arises from the long time limit of gravitational
collapse. Indeed, if a static or stationary black hole is known to
arise as the endpoint of a dynamical evolution, its stability is not
seriously in question. We will consider an argument for stability
based on the second approach; however, it should be emphasized that
such an approach is physical, and is not a substitute for a complete
mathematical analysis of the first type.

To begin, let us observe that the stress--energy--momentum tensor
(\ref{aset2}) can be rewritten as
\be
T_{ab} = \rho(r) (v_aw_b + v_b w_a)
+ P(r)(g_{ab} + v_aw_b + v_bw_a)\ ,
\label{nullf}
\ee
where the null vectors $v_a$ and $w_a$ are defined by
\be
U_a = {1\over \sqrt{2}}(v_a + w_a)\ ,\ \ \ \ \
u_a = {1\over \sqrt{2}}(v_a - w_a) \ .
\ee
For our solutions $U_a = (\sqrt{f},-1/\sqrt{f},0,0)$ and $u_a =
(0,1/\sqrt{f},0,0)$ in the coordinates $(v,r,\theta,\phi)$, where
$v$ is an advanced time coordinate defined by $dv = dt + (1/f) dr$.

In \cite{vh} a stress tensor similar to (\ref{nullf}) was
used to obtain dynamical spherically symmetric collapse
solutions. That tensor is
\be
T_{ab} = {\dot{m}(r,v)\over 4\pi r^2}v_a v_b
 +  \rho(r,v) (v_a w_b + v_b w_a)
+ P(r,v) (g_{ab} + v_aw_b + v_bw_a)\ .
\label{dnullf}
\ee
The general solution of
the Einstein equations for this source, with the equation of state
$P=k\rho$, is
\be
ds^2 = -\left(1 - {2\bar{f}(v)\over r}+{2\bar{g}(v)\over (2k-1)r^{2k}}
\right)dv^2  + 2dvdr + r^2d\Omega^2 ,
\label{met(rv)}
\ee
where
\be
m(r,v) = \bar{f}(v) - {\bar{g}(v)\over (2k-1)r^{2k-1}}\ ,\ \ (k\ne 1/2),
\ee
\be
P(r,v) = k\ {\bar{g}(v)\over 4\pi r^{2k+2}} = k\rho(r,v) \ ,
\ee
and $\bar{f}(v)$ and $\bar{g}(v)$ are arbitrary functions. Now we can
choose the functions $\bar{f}$ and $\bar{g}$ such that
$\lim_{v\rightarrow \infty} \bar{f}(v)= M$ and $\lim_{v\rightarrow
\infty} \bar{g}(v)= Q^{2k}(2k-1)/2$. Then the metric (\ref{met(rv)})
at large values of advanced time coincides with the short hair black
hole (\ref{met}), (\ref{sh}). In this way we see that the static short
hair solutions result as the end point of gravitational collapse. The
metric (\ref{met(rv)}) is to the short hair black hole what the Vaidya
metric is to the Schwarzschild black hole.

We now elaborate on the connection between the
stress--energy--momentum tensors (\ref{nullf}) and (\ref{dnullf})
induced by the perturbations (\ref{Del}). Notice first that if the
functions $f, \rho$ and $P$ are all taken to have general $(r,v)$
dependence, the most general stress--energy--momentum tensor {\it must
be of the form} (\ref{dnullf}) \cite{vh}. Thus the perturbations
(\ref{Del}), when substituted into the tensor (\ref{nullf}), must lead
to a tensor of the general form (\ref{dnullf}); the extra $v_av_b$
term is effectively induced by the time dependence of $f$. Now since
we have the general solution (\ref{met(rv)}) with the source
(\ref{dnullf}), we  have implicitly the solution of the equations
governing the perturbations (\ref{Del}), to all orders in the
perturbations.  This is because the general solution (\ref{met(rv)})
can always be separated into a static part plus a time
dependent `perturbation'.  Thus, this argument shows stability
under perturbations of the type (\ref{Del}).

The most general spherical scalar perturbations are however not given
by (\ref{Del}). To obtain the most general perturbations of this type
we would need to add to the stress--energy--momentum tensor
(\ref{dnullf}) an `outgoing' $(r,v)$ dependent term proportional to
$w_aw_b$. Such a term comes effectively from including a metric
perturbation $\delta h(r,v)\ dvdr$. Although an exact collapse
solution with some matter outgoing to infinity is not known, one can
argue, by analogy with the numerical spherically symmetric scalar
field collapse \cite{chop}, that an outgoing component with positive
energy density will not affect stability when black hole formation
occurs.  In that system it is observed, in the so called
`supercritical' region of parameter space (where a black hole forms),
that an initially ingoing pulse oscillates radially as it collapses,
with some scalar waves escaping to infinity with each oscillation as
the collapse to a black hole proceeds.  Similar behaviour has been
observed numerically in the collapse of perfect fluids \cite{ae}, and
may occur for the type of anisotropic fluid discussed here.  While
this analogy is suggestive, we only have an argument for stability for
the spherical `ingoing' perturbations (\ref{Del}).


\section{Comment on a `no short hair' theorem}


As we have seen, some of the static spherically symmetric black hole
solutions (those for $k>1$) arising from gravitational coupling to
anisotropic media have hair---that is, parameters in the metric which
are not captured by surface integrals at spatial infinity. The energy
density falls to zero with radius $r$ faster than $1/r^4$. The
physical picture is that some matter protrudes from the black hole
event horizon without being pulled in, but does not `extend to
infinity' as in the electric field case. Matter can hover in a strong
gravitational field without collapsing completely only if its internal
pressures are sufficiently large. This appears to be the case for the
anisotropic fluid black holes with $k>1$.

More generally one can ask: How close to a black hole event horizon
can matter hover if the stress--energy--momentum tensor satisfies some
reasonable physical conditions? Or in other words, how `short' can
hair be?  This question has been posed, and there is a theorem based
on certain assumptions \cite{nqs}.  Specifically, the theorem states
that if matter is such that

\begin{list}{}
\item{(i)} the weak energy condition holds,
\item{(ii)} the energy density $\rho$ falls to zero faster than $r^{-4}$,
\item{(iii)} $T:= T_{ab}g^{ab} \le 0$,
\end{list}
\bigskip
\noindent then it follows that $Pr^4$ ($P=$ radial pressure) {\it is
negative and decreasing until at least 3/2 the radius of the event
horizon, beyond which this quantity increases to zero}. This behaviour
is interpreted as `long hair', and it is suggested \cite{nqs} that it
will occur for any non--linear theory coupled to gravity. The three
conditions above effectively imply that radial pressure is negative,
at least near the horizon.

We have presented black hole solutions that have arbitrarily short
hair for the parameter $k>1$. For our `short hair' black holes,
conditions (i) and (ii) hold, but condition (iii) does not:
$T=2\rho(k-1)$, which is positive for $k>1$.  Thus, our solutions do
not contradict the no short hair theorem because the condition $T\leq
0$ for it to hold is not satisfied. Thus, while this theorem does
apply to several cases as discussed in \cite{nqs}, it does not apply
to the anisotropic fluid.


\section{Discussion}


We have presented a class of black hole solutions arising from
Einstein gravity coupled to a type of anisotropic matter, and described
some of their properties. Perhaps their most unusual feature is the
possibility of arbitrarily short hair.

It is interesting to observe that these black holes can also be seen
as arising from a generalized electromagnetism. Consider the
stress--energy--momentum tensor
\be
 T_{ab}={1\over 4\pi}\ ( F_{ac} F_b^{\ c} -
{\alpha\over 4}\ g_{ab} F_{cd}F^{cd})\ ,
\label{gem}
\ee
where $F_{ab}$ is the field strength and $\alpha$
is a constant.\footnote{We do not know an action which leads to the
stress--energy--momentum tensor (\ref{gem}).} The field equation for
$F_{ab}$ is that obtained from $\nabla_a T^{ab}=0$. A solution of the
Einstein equations coupled to this matter is the metric (\ref{met})
and (\ref{sh}), with
\be
F_{ab} = \sqrt{{(2k-1)(k+1)\over 2}}\ {Q^k\over r^{k+1}}
\ (dt\wedge dr)_{ab}\ ,
\ee
and $\alpha = 2k/(k+1)$. We note the black hole solutions viewed as
arising from this matter source can be shown to be stable via an analysis
very similar to that given in \cite{CX} for the Reissner-Nordstrom
solution. Indeed, the only difference in the details are appropriate
replacements of $Q^2/r^2$ by $Q^{2k}/r^{2k}$.

It would be of interest to study the collapse of anisotropic fluids
numerically, especially in light of the black hole solutions we have
found. In particular, as has been done for the perfect fluid
\cite{ae}, one could seek a `critical' solution with a self--similarity
ansatz. If found, such a solution could be used as the starting point
of perturbation theory \cite{ae,kha} for finding critical exponents
associated with the parameters $M$ and $Q$.

\medskip

\noindent{\it Note added}: After completing this work we became aware
of related results by Magli and Kijowski \cite{MK}. These authors
consider equilibriuim, spherically symmetric stars composed of
anisotropic elastic matter. We thank G. Magli for bringing this to our
attention.

\section{Acknowledgements} V.H. thanks Ted Jacobson for email discussions.
His work was supported by NSF grant PHY-9514240 to the Pennsylvania
State University, and by a Natural Science and Engineering Research
Council (Canada) grant to the University of New Brunswick.


\begin{references}

\bibitem{israel} W.~Israel, Phys. Rev. {\bf 164}, 1776 (1967); Commun.
Math. Phys. {\bf 8}, 245 (1968).

\bibitem{carter} B. Carter, Phys. Rev. Lett. {\bf 26}, 331 (1971).

\bibitem{hawking} S. W. Hawking, Commun. Math. Phys. {\bf 25}, 152
(1972).

\bibitem{robinson} D. C. Robinson, Phys. Rev. Lett. {\bf 34}, 905 (1975).

\bibitem{mazur} P. O. Mazur, J. Phys. A {\bf 15}, 3173 (1982).

\bibitem{bunting} G. Bunting, (unpublished).

\bibitem{nqs} D.~Nunez, H.~Quevedo and D.~Sudarsky,  Phys. Rev. Lett.
{\bf  76}, 571 (1996).

\bibitem{dewitt} B.S.~DeWitt, in {\it Gravitation: An Introduction to
Current Research\/}, edited by L.~Witten (Wiley, New York, 1962);
Phys. Rev. {\bf 160}, 1113 (1967).

\bibitem{dd} J.D.~Brown and D.~Marolf, Phys. Rev. D {\bf 53},  (1996).

\bibitem{vh} V.~Husain, Phys. Rev. D {\bf 53}, R1759 (1996).

\bibitem{chop} M. Choptuik,  Phys. Rev. Lett. {\bf 70}, 9 (1993).

\bibitem{ae} A. Abraham and C. R. Evans, Phys. Rev. Lett.
{\bf 70}, 2980 (1993).

\bibitem{CX} S. Chandrashekar and B. C. Xanthopolous, Proc. Roy Soc.
{\bf A 367} 1 (1979); S. Chandrashekar, {\it The Mathematical Theory of
Black Holes}, (Clarendon Press, Oxford, 1983).

\bibitem{kha} T. Koike, T. Hara, and A. Adachi, Phys. Rev. Lett. {\bf 74},
5170 (1995).

\bibitem{MK} G. Magli and J. Kijowski, Gen. Rel. Grav. {\bf 24}, 139 (1992).

\end{references}
\end{document}